\documentclass[aip, pop, amsfonts, floatfix, amssymb, amsmath ,reprint ,subscriptadress]{revtex4-1}
\usepackage{times}
\usepackage{bm}
\usepackage{hyperref}
\usepackage{graphicx}
\usepackage{graphics}
\usepackage{epsfig} 
\usepackage{epstopdf}
\usepackage{color}

\begin{document}

\title{Gyrokinetic studies of core  turbulence features in ASDEX Upgrade H-mode plasmas}
%
\author{A.~Ba\~n\'on Navarro}
\email{abanonna@ipp.mpg.de}
\affiliation{Max-Planck-Institut f\"ur Plasmaphysik, Boltzmannstrase 2,  85748 Garching, Germany}
\author{T.~Happel}
\affiliation{Max-Planck-Institut f\"ur Plasmaphysik, Boltzmannstrase 2,  85748 Garching, Germany}
\author{T.~G\"orler}
\affiliation{Max-Planck-Institut f\"ur Plasmaphysik, Boltzmannstrase 2,  85748 Garching, Germany}
\author{F.~Jenko}
\affiliation{Max-Planck-Institut f\"ur Plasmaphysik, Boltzmannstrase 2,  85748 Garching, Germany}
\affiliation{Max-Planck/Princeton Center for Plasma Physics}
\affiliation{Department of Physics and Astronomy, University of California, Los Angeles, California 90095}
\author{J.~Abiteboul}
\affiliation{Max-Planck-Institut f\"ur Plasmaphysik, Boltzmannstrase 2,  85748 Garching, Germany}
\author{A.~Bustos}
\affiliation{Max-Planck-Institut f\"ur Plasmaphysik, Boltzmannstrase 2,  85748 Garching, Germany}
\author{H.~Doerk}
\affiliation{Max-Planck-Institut f\"ur Plasmaphysik, Boltzmannstrase 2,  85748 Garching, Germany}
\author{D.~Told}
\affiliation{Max-Planck-Institut f\"ur Plasmaphysik, Boltzmannstrase 2,  85748 Garching, Germany}
\author{the ASDEX Upgrade Team}
\affiliation{Max-Planck-Institut f\"ur Plasmaphysik, Boltzmannstrase 2,  85748 Garching, Germany}

\date{\today}

\begin{abstract}
 Gyrokinetic validation studies are crucial in developing confidence in  the model 
incorporated in numerical simulations and thus improving their predictive capabilities. 
As one step in this direction, we  simulate an ASDEX Upgrade discharge with the GENE code, and 
analyze various fluctuating quantities and compare them to experimental measurements. 
The approach taken  is the following.  First, linear simulations are  performed in order  to determine  the turbulence regime.
  Second, the heat fluxes in nonlinear simulations are   matched  to experimental fluxes by varying the logarithmic ion temperature gradient within the expected   experimental error bars.  Finally, the dependence of various quantities with respect to the ion temperature gradient is analyzed in  detail. It is found that density and temperature fluctuations can vary significantly with small changes in this parameter, thus   making  comparisons  with experiments very sensitive to  uncertainties in the experimental profiles.  However,  cross-phases are more robust, indicating     that they  are   better  observables  for comparisons between  gyrokinetic simulations and experimental measurements.
\end{abstract}
\maketitle


\section{Introduction}


It has been known for several decades that energy and particle confinement  in tokamak  plasmas  
are mainly degraded by turbulence driven by steep temperature and density gradients.
For this reason, the characterization  and understanding of these turbulent  processes is a very important task  in order to improve the performance of present  experiments as well as future fusion reactors.

Due to the strong background magnetic field and the low collisionality   in tokamak plasmas,  gyrokinetic theory has been established as the most appropriate theoretical framework for the study of turbulent transport in the plasma core.  
In order to improve  confidence in the numerical results obtained with gyrokinetics  and to establish a solid understanding of turbulent transport across the whole range of plasma parameters, it is  very important to perform  direct comparisons between  simulations and experimental measurements.  In this respect, with the recent development and improvements in fluctuations diagnostic,
it is now possible to measure turbulence features with high precision, allowing  for quantitative comparisons between  experimental data and  results of nonlinear gyrokinetic simulations.   These validation studies are crucial in developing confidence in  the models and improving the predictive capabilities of the numerical simulations.

 In a recent paper~\cite{tim14}, we have compared  density fluctuation levels measured with a new Doppler reflectometer  installed in ASDEX Upgrade and simulation results obtained with the gyrokinetic GENE code.
 We  extend the previous work by additionally presenting  simulation results of  density wavenumber spectra, electron temperature fluctuation levels as well as  cross-phases between different  quantities.  One of the reasons for analyzing 
 electron temperature fluctuation levels and cross-phases is that a new Correlation Electron Cyclotron Emission (CECE) system
 is expected to be installed and to be in operation in 2015 in ASDEX Upgrade. Therefore, the gyrokinetic results presented in this paper 
 can provide guidance for the on-going development of the diagnostic.  From a more fundamental point of view, we will also  investigate the variation of these quantities  with respect to  various physical  input parameters. This information can  be used to characterize core turbulence features in ASDEX Upgrade  plasmas.

The paper is organized as follows. In Section~\ref{discharge}, an overview of the chosen plasma discharge  analyzed  is given. A description of the  gyrokinetic simulation method used is described  in  Section~\ref{overview}. Micro-instability studies from linear gyrokinetic simulations are outlined in Section~\ref{linear}. The main results of the paper are shown in Section~\ref{nonlinear}. Core turbulence features such as heat fluxes, density fluctuation amplitudes and  spectra, temperature fluctuation amplitudes as well as  cross-phases between these quantities will be presented in  detail, followed by a discussion in Section~\ref{discussion}.  Finally, conclusions and  future work will be discussed in Section~\ref{conclusions}.  

\section{Overview of the plasma discharge  \label{discharge}}
\begin{figure}[!ht]
\resizebox{\columnwidth}{!}{
\begin{tabular}{c}
\includegraphics[]{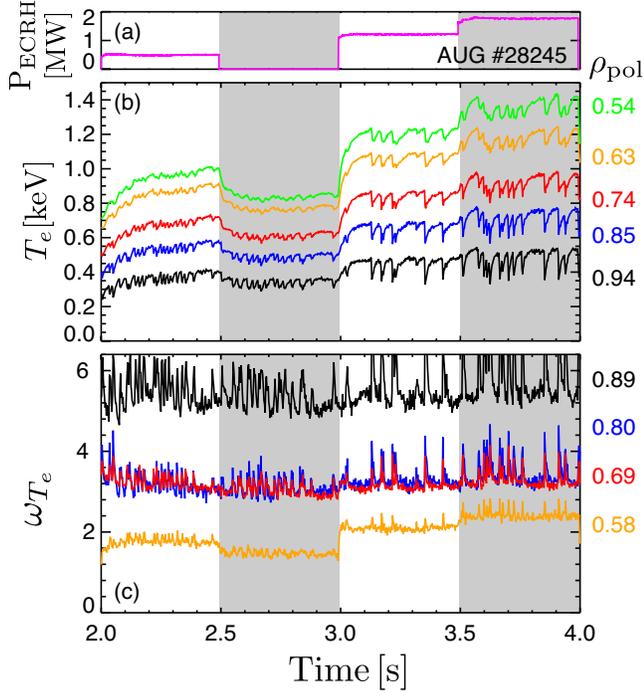}
\end{tabular}
}
\caption{(color online) Time traces of  ASDEX Upgrade discharge $\#28245$. (a) ECRH heating power,  (b) electron temperature and (c) logarithmic  electron temperature gradient at different radial positions. The data is analyzed in the time intervals shaded in grey.}
\label{fig_exp}
\end{figure}

The ASDEX Upgrade discharge  $\#28245$ analyzed in this paper was operated in the high-confinement regime (H-mode). It was planned   to study the turbulence characteristics in both the ion temperature gradient (ITG) and trapped electron mode (TEM) regimes, through a transition from one regime to another.  This transition can be achieved by modifying locally  the electron temperature gradient, which affects the TEM instability.   This is obtained by changing   the electron cyclotron resonance heating power (ECRH) by steps during the discharge.

An overview of several relevant time traces is given in Fig.~\ref{fig_exp}.
The ECRH power $P_{\rm{ECRH}}$ is deposited at  $\rho_{\rm{pol}} = 0.5$ from $2.0 -4.0$ seconds.  
Here,  $\rho_{\rm pol}$ is the normalized poloidal flux radius. 
 At this heating location, $P_{\rm{ECRH}}$ is varied subsequently between $0.5, 0.0, 1.2$ and $1.8 \,\rm MW$ (a).  The influence of the stepped heating power can be clearly observed in the electron temperature $T_e$ (b). Finally,  the strongest increase of the logarithmic temperature gradient  $\omega_{T_e}$ is observed at $\rho_{\rm pol} \approx 0.6$ when an ECRH power of $1.8$ MW is applied (c). For this reason, this case and the one without ECRH will be analyzed in detail,  as  examples  of the two extreme cases. They correspond to the time windows shaded in grey in  Fig.~\ref{fig_exp}. Within these time windows, we  have  simulated  three different radial locations:  $\rho_{\rm pol} = \{0.6, 0.7, 0.8 \}$, making a total of six different  scenarios to be studied with  gyrokinetic simulations. 

The physical parameters for each of these six cases are given in Table~\ref{table_exp}. There, the reference length is defined  as $L_{\rm ref}  = \sqrt{\Psi_{\rm tor, sep} / \pi B_{\rm ref}}$, where $\Psi_{\rm tor, sep}$ is the toroidal flux at the separatrix and $B_{\rm ref}$ is the magnetic field on axis.  Typically, this reference length is comparable but not identical to the tokamak minor radius.  The logarithmic gradients are defined as in Ref.~\onlinecite{told13}: $\omega_{X} = -\frac{1}{X}\frac{{\rm d}X}{{\rm d} \rho_{\rm tor}}$ with $X \in \{T_i, T_e, n_e  \}$, and $\rho_{\rm tor}$ the normalized toroidal flux radius. The magnetic shear is given as $\hat{s}  = \frac{\rho_{\rm tor}}{q}\frac{{\rm d} q}{{\rm d} \rho_{\rm tor}}$, where $q$ is the safety factor, and the electron beta is defined as $\beta_{e} = 2 \mu_0 n_e T_e / B_{\rm ref}^2$. Here, $T_i$ and $T_e$ are the ion and electron temperature, respectively,  and $n_e$ is the electron density.
\begin{table}[!h]
\resizebox{\columnwidth}{!}{
\begin{tabular}{c|c|c|c|c|c|c}
   \hline \hline
  Time [s] & 2.65-2.95 & 2.65-2.95 & 2.65-2.95 & 3.65-3.95 & 3.65-3.95 & 3.65-3.95 \\ \hline
  $\rho_{\rm pol}$ & 0.60 & 0.70 & 0.80 & 0.60 & 0.70 & 0.80 \\ 
  $\rho_{\rm tor}$ & 0.47 & 0.56 & 0.67 & 0.47 & 0.56 & 0.67 \\ 
  $\hat{s}$ & 0.73 & 1.05 & 1.47 & 0.59 & 1.00 & 1.63 \\ 
  $q$ & 2.39 & 2.82 & 3.56 & 2.22 & 2.57 & 3.24 \\ 
  $\omega_{T_i}$ & 1.69 & 2.05 & 2.01 & 1.10 & 1.49 & 2.06 \\ 
  $\omega_{T_e}$ & 1.44 & 2.06 & 1.82 & 2.15 & 2.11 & 2.16 \\ 
  $\omega_{n_e}$ & 0.01 & 0.24 & 0.48 & 0.59 & 0.43 &  0.49 \\ 
  $\beta_e \,[\%] $ & 0.29& 0.24& 0.19 & 0.50 & 0.39& 0.29 \\ 
  $T_i  \,[\rm keV]$ & 0.77 & 0.64 & 0.50 & 0.86 & 0.76 & 0.64 \\ 
  $T_e \, [\rm keV]$ & 0.80 & 0.68 & 0.54 & 1.23 & 1.00 & 0.80 \\ 
  $n_e \, [10^{19} \, {\rm m}^{-3}]$ & 4.40 & 4.36 & 4.25 & 4.83 & 4.63 & 4.39 \\ 
  $R_{\rm axis}/L_{\rm ref}$ & 2.51 & 2.51 & 2.51 & 2.59 & 2.59 & 2.59 \\ 
  $B_{\rm ref} \, [{\rm T}]$ & 2.21 & 2.21 & 2.21 & 2.19 & 2.19 & 2.19 \\ 
  $L_{\rm ref} \, [{\rm m}]$ & 0.68 & 0.68 & 0.68 & 0.66 & 0.66 & 0.66 \\ 
  $\rho_{s} \, [{\rm cm}]$    & 0.18  &  0.17   &  0.15      &  0.23       &   0.21  & 0.18 \\
   \hline \hline
\end{tabular}
}
\caption{Physical parameters for the six simulated ASDEX Upgrade cases. $R_{\rm axis}$ is the major radius at the magnetic axis and $\rho_{s} = c_s /  \Omega_{i}$ is a reference gyroradius defined with the ion sound speed $c_s = \sqrt{T_e / m_i}$ and the ion gyrofrequency $\Omega_{i}$.}
\label{table_exp}
\end{table}
%
  

\section{Overview of the gyrokinetic simulation method \label{overview}}

The turbulence data  obtained in this paper are produced with the gyrokinetic code GENE,
which  solves self-consistently the  gyrokinetic-Maxwell system of  equations on a fixed grid in five dimensional phase space (plus time): two velocity coordinates $(v_{\parallel}, \mu)$ and three field-aligned coordinates  $(x,y,z)$. Here, $z$ is the coordinate along the magnetic field line, while the radial coordinate $x$ and the binormal coordinate $y$ are orthogonal to the equilibrium magnetic field. The velocity coordinates are, respectively, the velocity parallel to the magnetic field and the magnetic moment.  GENE has the possibility to  simulate  either a flux-tube  (local simulations) or   a full torus  (global simulations). In the former option,  it  is assumed that the relevant turbulent structures are  small with respect to the radial variation of the background profiles and gradients.  This allows to use periodic boundary conditions and thus the coordinates perpendicular to the magnetic field are Fourier transformed $(x,y) \rightarrow (k_x, k_y)$. For this work,  only the local version of the code has been employed.

The GENE code  is physically quite comprehensive and includes many features  (see Ref.~\onlinecite{gene} for  more details). For the ASDEX Upgrade scenario studied in this paper,  the following features of the GENE code were used: two particle species (deuterons and electrons), electromagnetic effects by solving the parallel component of   Amp\`ere's law,  external $E \times B$ shear,  parallel  flow shear  and a linearized Landau-Boltzmann collision operator with energy and momentum conserving terms~\cite{hauke13}.  
 Unless stated otherwise, the magnetic equilibrium geometry is taken from the TRACER-EFIT interface~\cite{pablos09}. Additionally,  GyroLES techniques have been used to reduce the accumulation of energy at the smallest scales (see Refs.~[\onlinecite{morel11,morel13,banon14}]). Further simulation details, such as resolution grid, box sizes, etc., are given in the following sections. 


\section{Micro-instability studies: Linear gyrokinetic simulations \label{linear}}

In order to calculate    turbulent transport fluxes, density and temperature fluctuation amplitudes, etc.,  nonlinear gyrokinetic  simulations are necessary. Nevertheless,  linear gyrokinetic simulations can provide  useful insights. For instance, they may allow us to identify the  underlying micro-instabilities  which drive  the turbulence  present in the experiments. They can also be used for convergence studies and,  since they are usually  computationally  cheap,  they can also be used  to  do scans in  different physical parameters.

\subsubsection{Nominal parameter set}

In linear simulations we calculate the growth rate and frequency of the most unstable mode present in the system for a given binormal wave vector $k_y$ and $k_x=0$. In this paper we choose to present $k_y$ in $\rm {cm}^{-1}$   instead of  the more common   $k_y \rho_s$ units.  This has been done in order to compare to experimental results. For reference,   $\rho_s$ values are given in Table~\ref{table_exp}. 

\begin{figure}[!h]
\resizebox{\columnwidth}{!}{
\begin{tabular}{c}
\includegraphics[]{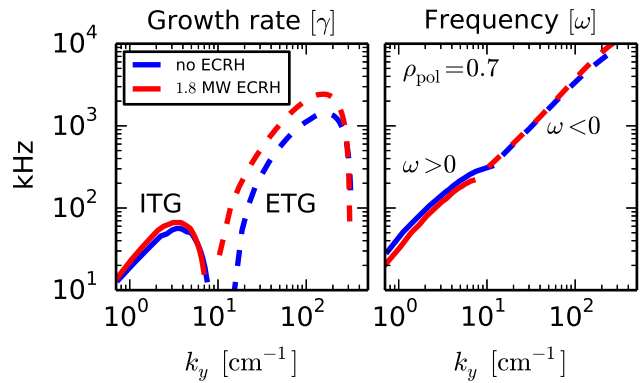} 
\end{tabular}
}
\caption{(color online) Linear growth rates (left) and frequencies (right) for the cases at $\rho_{\rm pol = 0.7}$ versus the binormal wavenumber $k_y$. The dashed lines in the figure indicate  negative frequency values.  Positive frequencies refer to modes drifting in the ion diamagnetic direction  and negative in the electron diamagnetic direction.    For these cases, they correspond to  ITG  and ETG modes, respectively.}
\label{fig_nom_par}
\end{figure}

 In Fig~\ref{fig_nom_par},  we display the growth rates ($\gamma$) and frequencies $(\omega)$ for 
the cases at $\rho_{\rm pol} =0.6$, because  similar conclusions are obtained for the other cases. In the figure, the negative frequencies are represented by dashed lines. 
The grid resolution was $\{ x, z, v_{\parallel}, \mu \} = \{ 31,32,48,16\}$. Convergence tests were performed at  higher resolutions and confirm the validity of the results.  
Several observations can be made. First, for low wavenumbers, 
ITG is the dominant instability. This is indicated by a positive frequency, which with the present normalization represents a frequency in the ion diamagnetic direction.  Second, for all the cases analyzed, the TEM mode is stable (studied with an eigenvalue solver). In fact, as  was already shown in Ref.~\onlinecite{tim14}, only with the combination of a  much higher electron temperature gradient  and a lower ion temperature gradient  than the ones measured  experimentally,   do TEM modes become unstable~\cite{normaliz}. Third, for higher wavenumbers, ETG is the dominant instability (indicated by a negative frequency).  
Finally,  there is practically no effect of the  ECRH on the ITG growth rates.   The main effect of ECRH is to increase the growth rates of ETG modes,  possibly leading to a subsequent  increase of  the electron heat flux for these cases.  However, in both cases, we expect their contribution to be small with respect to the ITG contribution. Moreover,    ETG modes  are not expected to influence  density and temperature fluctuations at low wavenumbers, which are dominated by ITG. Since  these are the scales 
measured by the diagnostics we are considering here, in this work we will limit ourselves to wavenumbers up to $k_{y}   = 10 \, {\rm cm^{-1}}$, thus excluding ETG modes, but allowing for  a significant reduction  in computational resources.
%

\subsubsection{Sensitivity studies with respect to the main physical parameters}

Physical  parameters  such as temperature, density,  magnetic equilibrium profiles, etc.,  are measured with experimental uncertainties. Since these values are used as input in the gyrokinetic codes, the uncertainty in these quantities could affect the simulation results. Therefore,   sensitivity studies are carried out, with the aim of  studying the effect of the different uncertainties on the simulation results. Ideally, these studies should be done in nonlinear gyrokinetic simulations. However, due to the  expensive computational effort associated, this is  in practice  unfeasible.  For this reason, this sensitivity study is  done   within linear gyrokinetic simulations.

\begin{figure}[!h]
\resizebox{\columnwidth}{!}{
\begin{tabular}{c}
\includegraphics[]{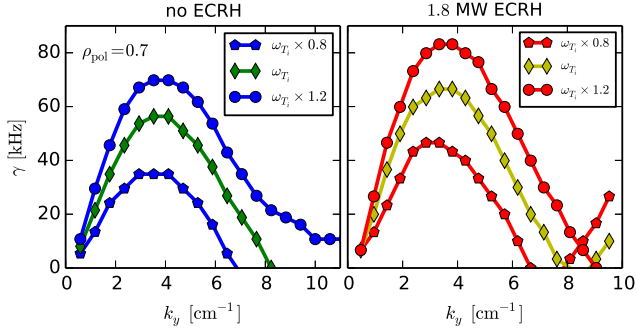}
\end{tabular} 
}
\caption{(color online) Linear growth rates  versus the binormal wavenumber $k_y$ for the cases at $\rho_{\rm pol}= 0.7$  with respect to the variation of the logarithmic ion temperature gradient.}
\label{fig_lin_var}
\end{figure}

 We have studied the sensitivity of the linear growth rate with respect to a variation of 
 $\pm 20 \, \%$  in the nominal value for different physical parameters,  such  as:  logarithmic ion temperature gradient ($\omega_{T_i}$), logarithmic electron temperature gradient ($\omega_{Te}$), logarithmic electron density gradient ($\omega_{n_e}$), electron to ion temperature ratio ($T_e /T_i$), collisionality ($\nu_{\rm col}$),  safety factor ($q$) and magnetic shear ($\hat{s}$). The main results are  summarized in Table~\ref{tab_lin_var}. For simplicity, we show only   cases at $\rho_{\rm pol} = 0.7$,  although similar conclusions were obtained  to the other radial positions.  The sensitivity studies for  $q$ and for $\hat{s}$ have been done using a  Miller-type magnetic equilibrium~\cite{miller}. As  is shown in Table~\ref{tab_lin_var}, the peak of the  growth rates are practically insensitive with respect to changes of $\pm 20 \%$ in $\omega_n$, $\omega_{Te}$, $\nu_{\rm col}$, $q$ and  $\hat{s}$. Changes in the peak of the  growth rate up to $20 \,\%$   are found for  $T_e /T_i$ variations. 
 The most critical parameter is $\omega_{T_i}$, whose $\pm 20 \%$ variation modifies the growth rate by up to $40 \%$ (see Fig.~\ref{fig_lin_var}). Based on these results, one could expect that  the   uncertainties in $\omega_{T_i}$  will have the largest influence in nonlinear gyrokinetic simulations. In the following, we will mainly focus on the influence of this parameter on nonlinear simulations. The influence of the $E \times B$ shear is in general  also expected to have a relevant impact on the transport in nonlinear simulations. However, the low value of the $E \times B$  shear for this particular discharge is such that its influence can safely be neglected. 
\begin{table}[!h]
\begin{tabular}{  | c c | c c |}
   \hline \hline
    No ECRH   &   $\Delta \gamma$&  $1.8$ MW ECRH & $\Delta \gamma$  \\ \hline \hline 
      $\omega_{Ti} \times \, 0.8$        & -38 \%  &      $\omega_{Ti} \times \, 0.8$        &      -30\%       \\ \hline
     $\omega_{Ti} \times \, 1.2$       &    +25 \%   &     $\omega_{Ti} \times \, 1.2$        &       +26\%   \\ \hline \hline
      $\omega_{T_e} \times \, 0.8$        &   -4 \%  &     $\omega_{T_e} \times \, 0.8$        &     -5\%    \\ \hline        
      $\omega_{T_e} \times \, 1.2$        &   +5 \%   &     $\omega_{T_e} \times \, 1.2$         &    +5\%   \\ \hline \hline
      $\omega_{n_e} \times \, 0.8$        &    +0  \% &      $\omega_{n_e} \times \, 0.8$       &   -5\%         \\ \hline  
      $\omega_{n_e} \times \, 1.2$       &  +0 \%    &   $\omega_{n_e} \times \, 1.2$        &   +6 \%     \\ \hline \hline
      $T_e / T_i \times \, 0.8$        &    +14 \%   &   $T_e / T_i \times \, 0.8$       &     +11 \%  \\ \hline
      $T_e / T_i \times \, 1.2$         &  -18 \%   &    $T_e / T_i \times \, 1.2$        &  -9\%  \\ \hline \hline
      $q \times \, 0.8$       &    -14 \%     &   $q \times \, 0.8$      &     -9 \%   \\ \hline
      $q \times \, 1.2$         &  +5 \%   &    $q \times \, 1.2$         &   +6\%   \\ \hline \hline
      $\hat{s} \times \, 0.8$        &    +0  \%         &     $\hat{s} \times \, 0.8$        &   -5\%   \\ \hline
      $\hat{s} \times \, 1.2$        &-4 \%     &      $\hat{s} \times \, 1.2$       &   +5 \%  \\ \hline \hline
      $\nu_{\rm col} \times \, 0.8$        &    +5 \%         &     $\nu_{\rm col} \times \, 0.8$        &   +6\%   \\ \hline
      $\nu_{\rm col} \times \, 1.2$        &  -4 \%     &     $\nu_{\rm col} \times \, 1.2$       &  -5 \%  \\ \hline \hline
   \end{tabular}
\caption{Percentage difference in maximum growth rate with respect to the nominal values for cases at $\rho_{\rm pol} = 0.7$  for various  parameters ($\omega_{Ti}$, \, $\omega_{Te}$, \, $\omega_{n}$, \, $T_{i} / T_{e}$, \, $q$, $\hat{s}$, and $\nu_{\rm col}$).}
\label{tab_lin_var}
\end{table}
%


\section{Core turbulence features: nonlinear gyrokinetic simulations \label{nonlinear}}

In order  to predict and compare with experimental results, nonlinear simulations are required. For the selected discharge, the grid resolution needed is  $\{ 256 \times 128 \times 32 \times  48 \times 16 \} $ points in   $\{ x, y, z, v_{\parallel}, \mu \}$ coordinates. A convergence test of the results with that resolution has been performed by comparison with nonlinear simulations with a double resolution in the perpendicular directions for a few  cases. Moreover, perpendicular box sizes  have been chosen in such a way that several correlation lengths fit in the box and convergence checks have also been done on this respect to ensure the validity of this choice.  Results of the simulations are  time-averaged over a range 
well exceeding the correlation time of the underlying turbulence.


\subsection{Turbulence ion and electron heat fluxes}

\begin{figure}[!h]
\resizebox{\columnwidth}{!}{
\begin{tabular}{c}
\includegraphics[]{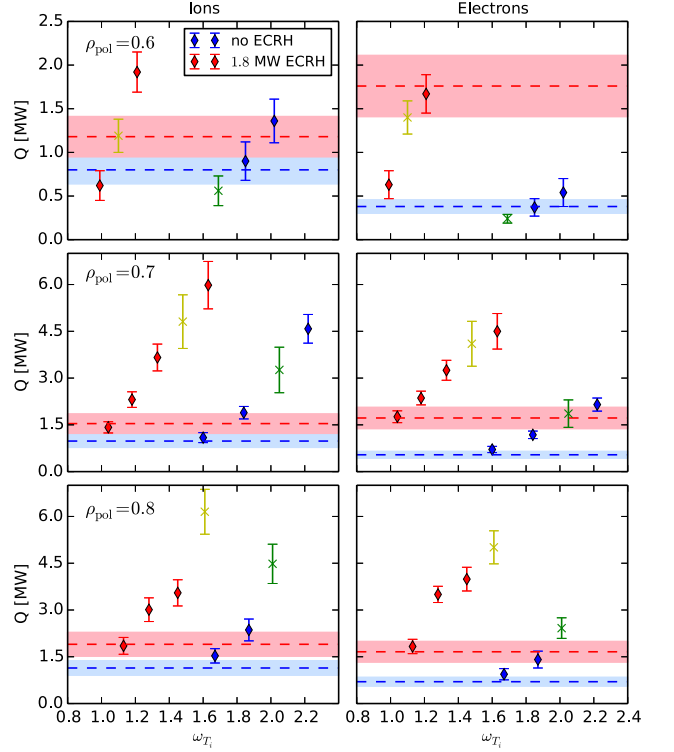} 
\end{tabular}
}
\caption{(color online) Comparison of experimental heat fluxes (dashed-lines) with those obtained from gyrokinetic nonlinear GENE simulations (markers) using the nominal parameters and the variation with respect to logarithmic  ion temperature gradient for all cases.  Ion heat fluxes (left) and electron heat fluxes (right) are depicted. The rows represent the radial positions $\rho_{\rm pol} = 0.6, 0.7, 0.8$. 
 The experimental values are obtained through  power balance analysis with the ASTRA code and the shaded regions are used to indicate the uncertainty of the  ASTRA values. In this case, a $20 \%$ error is assumed for all cases. In blue are the discharge parameters without ECRH heating and in red with ECRH heating. The nominal parameter  in each case is colored differently to distinguish it from the rest.}
\label{fig_stiff}
\end{figure}

In this section, we compare the experimental ion and electron heat fluxes obtained through power balance analysis with the ASTRA code  and the results from  nonlinear GENE simulations. As was done previously, we have grouped in the same plot the cases  taken at the same radial position.  The  results are shown in Fig.~\ref{fig_stiff}. 

\begin{figure}[!h]
\resizebox{\columnwidth}{!}{
\begin{tabular}{c}
\includegraphics[]{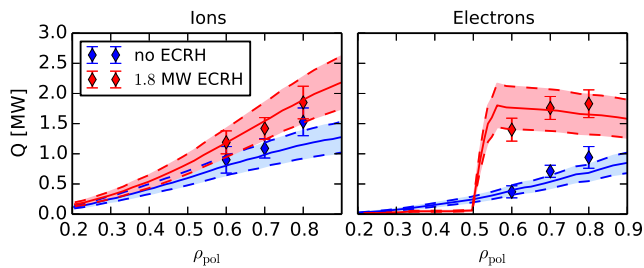} 
\end{tabular}
}
\caption{(color online) Comparison of experimental heat fluxes (dashed-lines) with those obtained from gyrokinetic nonlinear GENE simulations (markers) using the flux-matched  simulations.}
\label{fig_heat_comp}
\end{figure}

Ion (electron) heat fluxes are shown in the left (right) columns in Fig.~\ref{fig_stiff}. The rows represent radial positions $\rho_{\rm pol} = 0.6, 0.7, 0.8$. The dashed lines indicate the ASTRA results and their shaded regions are used to indicate the uncertainty of the  ASTRA values, where a $20 \ \%$ uncertainty  is assumed~\cite{told13}.
Based on the linear sensitivity studies, for each case, several simulations were performed  varying the logarithmic ion temperature gradient in steps of $\pm 10 \%$ with respect to the nominal values,  up to  a maximum of $\pm 30 \%$ variation. GENE simulation results are represented  by  the markers in the figure.  The cases without ECRH are colored in blue and with ECRH in red and for each set, the simulations with the nominal parameters  are colored  differently  (in  green for the cases without ECRH and in yellow for the cases with ECRH). The statistical error bar is an estimation of the standard deviation of the set of means of consecutive temporal sub-domains of the saturated state. Several conclusions can be obtained from Fig.~\ref{fig_stiff}. For the nominal parameters, the cases with ECRH produce more  ion and electron heat flux than the ones without ECRH for all positions.  Moreover, at   $\rho_{\rm pol} =0.6$, the ion heat fluxes match the experimental values without having to  vary the  gradient with respect to the nominal value. At this position,  only the electron temperature must be increased by   $10 \%$. However,  for the other positions, the values of the heat fluxes  obtained with the nominal parameters clearly overestimate  the  heat fluxes obtained with ASTRA by a factor of $2-3$.  We need to decrease $\omega_{T_i}$ by $20 \%$ in order to match the experimental results for the cases without ECRH. Whilst, for the  cases with ECRH heating, it has to be decreased by a maximum of $30 \%$.  In Fig.~\ref{fig_heat_comp}, only  the flux-matched results are compared with the experimental heat fluxes.  We can conclude that agreement of the transport levels within the errors bars can be achieved, since   even small  uncertainties in the temperature profile itself may translate to relatively large ones (up to $ 20 - 30 \%$) in the logarithmic gradients~\cite{tobias14, tim14}.    
%


\subsubsection{Turbulence electron density   amplitudes}

Since turbulent fluxes are caused by plasma fluctuations on microscopic scales, it is necessary to validate gyrokinetic codes on a microscopic level. In this regard, a new Doppler reflectometer has been  recently  installed in ASDEX Upgrade (see Ref~\onlinecite{tim14} for more details on the diagnostic), which is able to measure electron density fluctuation amplitudes ($\tilde{n}_e$).  In order to compare experimental and numerical results, a synthetic diagnostic must be implemented in GENE to reproduce the measurement process of the reflectometer.  Two kinds of synthetic diagnostics can be employed.  A first approach  consists in simply filtering the data in order to taking  into account only the location  and the  wavenumbers that the diagnostic  measures. This translates  to take into account only fluctuations at the outboard mid-plane ($z=0$),  averaged over a finite radial length  and then to select   the range of  perpendicular wavenumber that are measured in each case ($k_{y, \rm min}^{\rm measured} \le k_y \le k_{y, \rm max}^{\rm measured}$ in GENE).   A  more sophisticated method   uses a full-wave code to simulate also the  incidence and reflection of the wave into the gyrokinetic  turbulent data~\cite{hillesheim12}. This work is in progress and only preliminary results are available with this synthetic diagnostic (see Ref.~\onlinecite{carsten}).  For this reason,  we  have only used   the filtering method for this work.

\begin{figure}[!h]
\resizebox{\columnwidth}{!}{
\begin{tabular}{c}
\includegraphics[]{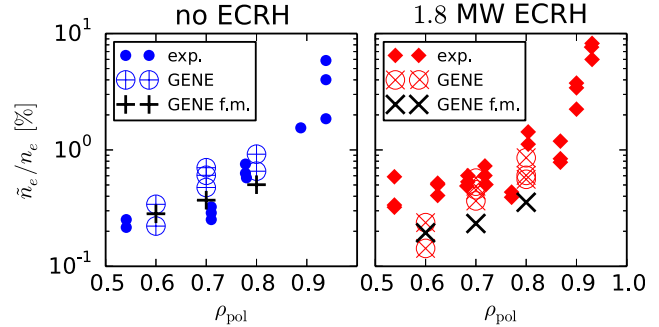} 
\end{tabular}
}
\caption{(color online) Electron density fluctuation amplitudes   at different radial positions. Blue-plus markers  represent the data without ECRH heating,  red-cross markers with ECRH and experimental results are in full circles.  The  flux-matched  (GENE f.m.)  simulations  are marked  differently.  Fluctuation data is analyzed  at the outboard mid-plane, averaged over a finite radial length  and  with  perpendicular wavenumbers between $4  \le k_y \,  \le  8 \, [{\rm cm}^{-1}]$.}
\label{fig_spec_dens}
\end{figure}

Comparison of the experimental and simulated $\tilde{n}_e$ are shown in Fig.~\ref{fig_spec_dens}.  The simulations that match the experimental ion  heat fluxes are  shown with a  different marker to  distinguish them from the rest. The fluctuation data is analyzed   considering  only perpendicular wavenumbers between $4  \le k_y \,  \le  8 \, [{\rm cm}^{-1}]$.  The experimental values are scaled by a common  factor since the  measurements are  in arbitrary units.  For this reason, only the  shape of radial turbulence level profiles and the effect of  ECRH can be used for comparison.  We decided to scale the experimental values   to try to match  the case without ECRH.  As is shown in  the left plot of Fig.~\ref{fig_spec_dens},  we obtain a  remarkable agreement between experimental and simulations results in the radial trend. For the case of $1.8$ MW ECRH, there is also  a good agreement in the turbulence level profile. However, with the scale used, the fluctuation levels are clearly underestimated with  respect to the ones measured experimentally. In particular, the flux-matched results present the biggest discrepancy with respect to the experimental measurements.   Finally, from this figure we can also observe how sensitive  the density fluctuations are with respect to  variations in the ion logarithmic temperature gradient. For instance, a $30 \%$ reduction in the  logarithmic gradient can reduce    density fluctuation  levels by  more than a factor of $2$.
%


\subsubsection{Turbulence electron density  spectra}
\begin{figure}[!h]
\resizebox{\columnwidth}{!}{
\begin{tabular}{c}
\includegraphics[]{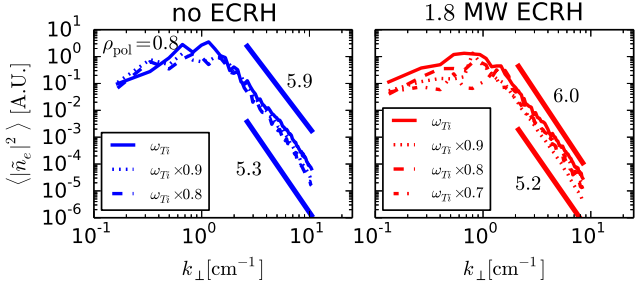} 
\end{tabular}
}
\caption{(color online) Electron density fluctuations spectra  at $\rho_{\rm pol} = 0.8$ including the variation of the logarithmic ion temperature gradient.  The solid lines represent the wavenumber range where the fit to a power law was done. Spectral indices  for the flattest and steepest spectra are also indicated in the figure for each case.}
\label{fig_spec_dens_1}
\end{figure}
\begin{figure}[!h]
\resizebox{\columnwidth}{!}{
\begin{tabular}{c}
\includegraphics[]{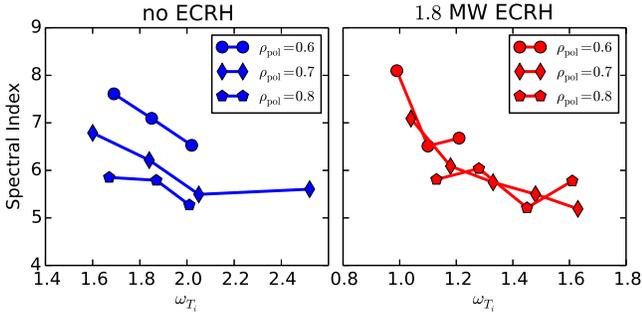} 
\end{tabular}
}
\caption{(color online) Calculated spectral indices  versus  the logarithmic ion  temperature gradient for different radial 
positions and ECRH scenarios. A decrease of the  spectral indices with respect to the ion temperature gradient
 is observed for most of the  cases.}
\label{fig_spec_dens_2}
\end{figure}
\begin{figure}[!h]
\resizebox{\columnwidth}{!}{
\begin{tabular}{c}
\includegraphics[]{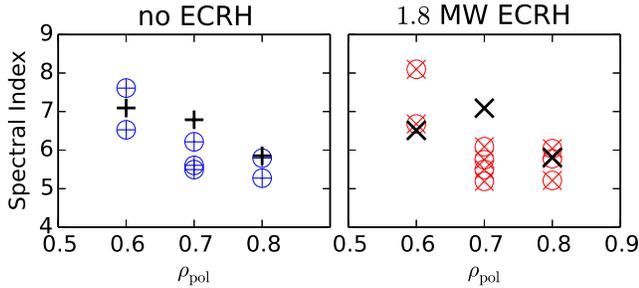} 
\end{tabular}
}
\caption{(color online) Calculated spectral indices   versus the radial  position. 
Blue-plus markers  represent the data without ECRH heating and red-cross markers with ECRH. 
The flux-match simulations are marked  differently. A decrease of the  spectral indices  with respect to the radial position is observed. }
\label{fig_spec_dens_3}
\end{figure}

The knowledge of the power-law spectra of a physical quantity is important for the understanding of the underlying physics and useful for providing constraints for simple physical models.  Based  on Kolmogorov-type arguments~\cite{k41}, turbulence is  generally associated with universal power-law spectra.  However, as shown in Refs~\onlinecite{vasil13,silvio14, banon14},   this is generally not the case in plasma turbulence, and  different  power-law  spectra indices can be found depending on the type of mechanism which drives  or dissipates energy in  the system.  Furthermore, the knowledge of  wavenumber spectra could be  important for a clear identification of the  turbulent regimes driven by different  microinstabilities and can be used to further validation of the gyrokinetic model.

The  results   of the   electron density fluctuation  spectra at $\rho_{\rm pol} = 0.8$  for different logarithmic ion temperature gradients are  shown in Fig.~\ref{fig_spec_dens_1}.    The solid lines represent the wavenumber range where a fit to a power law is shown: $\left <  | \tilde{n}_e |^2  \right> =a \, k_{\perp}^{-b}$, where $b$ is the spectral index, and $\left < \right> $ represents an average over all the coordinates except $k_{\perp}$. Spectral indices  for the flattest and steepest spectra are  indicated in the figure. The calculated spectral indices  are also shown in  Figs.~\ref{fig_spec_dens_2} and ~\ref{fig_spec_dens_3} with respect to  the ion temperature gradient and to the radial position, respectively.  These figures present a clear qualitative behavior: the spectral index decreases with the increase of turbulence drive. Additionally, the spectral indices also decrease when going from the inner   to the outer core position. Moreover, the magnitude of the exponents cover a wide range of values, approximately  from $4$ to $9$.
 Although  density fluctuation spectra were not  measured with the Doppler reflectometer for this discharge, a similar trend  has been also reported in Ref.~\onlinecite{carolin} for various ASDEX Upgrade discharges. 
 
 These  results   indicate that the     turbulence driven by ITG modes exhibits  non-universal power laws, whose  spectral indices  could depend on several physical parameters.   Future work in this respect will be to study also if TEM modes exhibit similar properties. If this was the case, then it  would  become  very  difficult to distinguish a type of instability by measuring only its characteristic spectral index.  


\subsection{Turbulence electron temperature  amplitudes} 

At the time the discharge was performed, no temperature fluctuation measurements  were available. However, a    Correlation Electron Cyclotron Emission (CECE) diagnostic  is currently   installed on ASDEX Upgrade, and electron temperature fluctuation profiles will be available in the future campaigns.

The CECE  diagnostic  measures perpendicular electron temperature fluctuations  ($\tilde{T}_{\perp,e}$)  in the  long wavelength range (relevant for ITG and TEM modes) and  is not sensitive short wavelengths (ETG modes).  This diagnostic  presents an inherent limitation in the lowest  fluctuation level that can be  detected.  This noise level  depends on the   physical parameters of the discharge,  but typical values are between $0.2-0.3 \, \%$~\cite{white08}.   

A detailed description of CECE modeling  in DIII-D is given in  Ref.~\onlinecite{shafer12}.  This synthetic diagnostic has already  been  implemented  in GENE for DIII-D discharges~\cite{tobias14} and a similar synthetic  diagnostic will be implemented for ASDEX Upgrade discharges. However,  this diagnostic could be not used in this work  since it requires the knowledge of the  CECE configuration during the discharge. For this reason, we have considered a simpler synthetic  diagnostic, which consists in  filtering the gyrokinetic data to  the positions and  wavenumbers   that are expected to be measured in ASDEX Upgrade.  Consequently, the gyrokinetic data analysis  results are restricted to  the outboard mid-plane position ($z=0$), averaged  over the finite radial length and summing  all perpendicular wavenumbers (since short  wavelengths are not simulated). In order to better model the actual diagnostic, one should also apply a filter in the frequency space. However, this filter  will also  depend on the specific range of frequencies measured. Since we do not have access to this information, we have considered all the frequencies in the analysis.  Therefore, the following results should  be  only   used as an  approximated  indication of the fluctuation  amplitudes  that could  be  detected  with CECE for this discharge.  Nevertheless, we do not expect  radial trends to change with respect to  a more  sophisticated synthetic diagnostic approach and  only the amplitudes are likely to be rescaled~\cite{holland09}.

\begin{figure}[!h]
\resizebox{\columnwidth}{!}{
\begin{tabular}{c}
\includegraphics[]{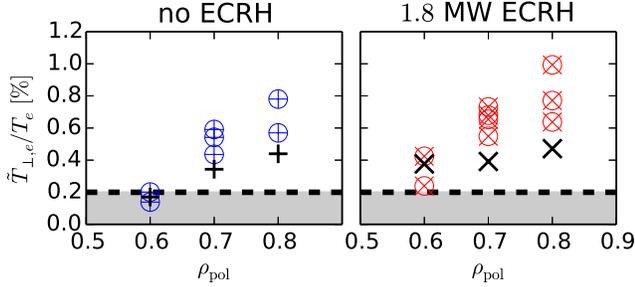} 
\end{tabular}
}
\caption{(color online) 
Percentages of   electron perpendicular  temperature  fluctuation  amplitudes   at different radial positions. Blue-plus  markers  represent the data without ECRH heating and red-cross markers with ECRH. The flux-match simulations  are marked differently.  The area shaded in grey  indicates a typical  noise level of the CECE diagnostic.}
\label{fig_fluc_temp}
\end{figure}

The main results are shown in Fig.~\ref{fig_fluc_temp}.
The perpendicular electron temperature fluctuation  amplitudes  go from a minimum of  $0.1 \,\% $ at the inner position  to  a maximum of  $1.0 \,\%$  at the outer core position. Therefore, assuming a noise-level in the order of $0.2 -0.3 \,\% $ for  the CECE diagnostic, this  could imply that only the fluctuations  at the outer  core  positions (starting from $\rho_{\rm pol} \geq 0.6$)  could be detected. As for the case of the density amplitudes, we also observe a large variation of the fluctuation amplitudes with respect to the changes in the logarithmic ion temperature gradient.

%


\subsection{Turbulence cross-phases} 
\begin{figure}[h]
\resizebox{\columnwidth}{!}{
\begin{tabular}{c}
\includegraphics[]{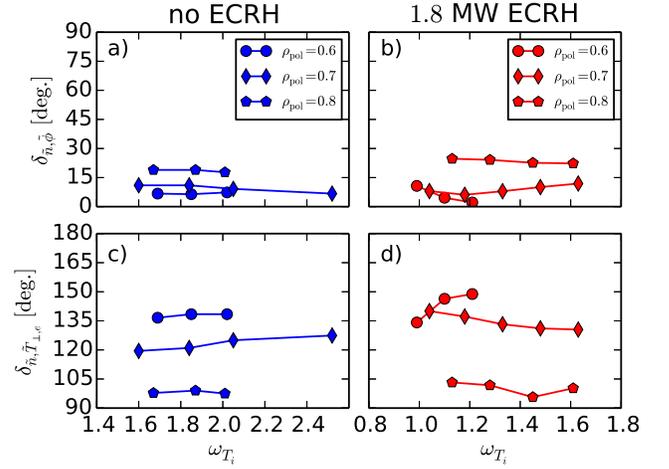} 
\end{tabular}
}
\caption{(color online)  Calculated  cross-phases for the different cases versus the logarithmic  ion  temperature gradient. Cross-phases between density and electrostatic potential fluctuations (a) and (b). 
Cross-phases between  electron density and  temperature fluctuations (c) and (d).  The cross-phases seem to be rather insensitive  with respect to  changes in the logarithmic ion temperature gradients.}
\label{fig_cross_2}
\end{figure}

Doppler reflectometers  can    be coupled  to the CECE diagnostics to calculate cross-phases between  electron density and  temperature fluctuations~\cite{hillesheim13}. This measurement  is important for gyrokinetic validation studies since it represents a  relationship between different fluctuating  quantities (density and temperature in this case). In addition,  this cross-phase could be  also related to the cross-phase that determines the turbulent heat fluxes (electrostatic potential  and temperature fluctuations).  
For this reason, in addition to the cross-phase that can be measured experimentally, we will also show the cross-phase between electrostatic potential and electron density fluctuations, so we can relate the cross-phases measured experimentally   to  the turbulent heat fluxes.  The cross-phases are here defined as   $\delta_{\tilde{n}, \tilde{T}_{\perp, e}} = \tan^{-1} (\Im (\tilde{n}/\tilde{T}_{\perp, e}) / \Re  (\tilde{n}/\tilde{T}_{\perp, e}))$, and, $\delta_{\tilde{n}, \tilde{\phi}}= \tan^{-1} (\Im (\tilde{n}/\tilde{\phi}) / \Re  (\tilde{n}/\tilde{\phi}))$.
 
Fig.~\ref{fig_cross_2} shows  the variation of the cross-phase versus the ion temperature gradient
integrated over binormal wavenumbers in the range of $1 \, {\rm cm}^{-1}  \le k_y \le  10 \, {\rm cm}^{-1} $. 
For most of the cases, the phases remain rather invariant with respect to the variation of this parameter. 
Therefore, this result seems to indicate that the cross-phase is a better observable to identify the type of instability which drives the turbulence in experiments, since  TEM instability is expected to  exhibit   different   cross-phases~\cite{dannert05}.   

\begin{figure}[h]
\resizebox{\columnwidth}{!}{
\begin{tabular}{c}
\includegraphics[]{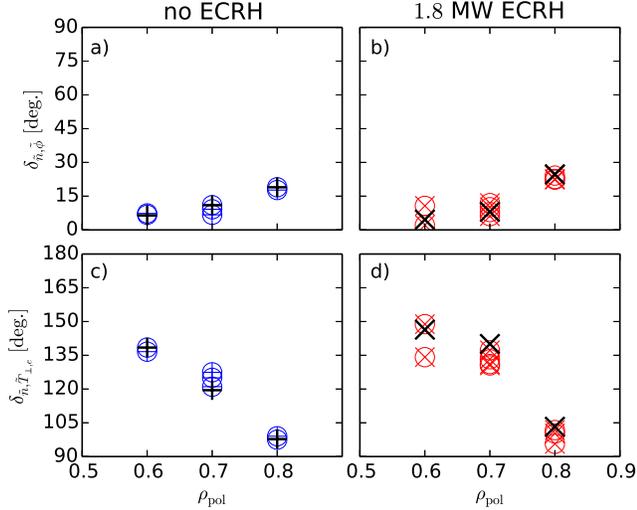} 
\end{tabular}
}
\caption{(color online) Calculated cross-phases  versus the  radial position. 
Cross-phases between density and electrostatic potential fluctuations (a) and (b).
Cross-phases between  electron density and  temperature fluctuations (c) and (d). 
The flux-match simulations  are marked differently.}
\label{fig_cross_3}
\end{figure}

In Fig.~\ref{fig_cross_3}, the cross-phase are displayed versus the radial positions.
Regarding the  cross-phase between density and electrostatic potential fluctuations (a) and (b),  we see 
that they are practically in phase, i.e.  close to $0$.  In addition, an  increase of the phase with  the radial position is also observed, going from practically $0$ degrees at $\rho_{\rm pol}  = 0.6$ to  approximate by $25$ degrees at the outer core position.  On the contrary, for  the cross-phase between electron density and temperature fluctuations (c) and (d), we see a decrease  with the radial position, going from around $150$ degrees in the inner position to a   value of  $90$ degrees, which  result in an increase   of the electron heat flux.   Similar values of this cross-phase   have also been measured in DIII-D and calculated  with GYRO  in Ref.~\onlinecite{white10}.  Furthermore, these values have also been found by GENE for these discharges, see Ref.~\onlinecite{tobias14}. These observations  could  be explained  in the following way.   In  the inner position, the population of the trapped particles  that contributes to the ITG instability is small,  so    the electrons behave almost adiabatically. Because of this,  density and potential fluctuations are  in phase  and  density and perpendicular temperature fluctuations are almost out of phase (i.e., close to $90$ degrees), thus producing  negligible electron heat flux. With increasing the radial position, the  population of trapped particles  increase  and a deviation   of the electron  adiabaticity is observed.
 For this reasons, both cross-phases  approach   to $90$ degrees, with the  subsequent   increase of electron heat flux.

\begin{figure}[htp]
\resizebox{\columnwidth}{!}{
\begin{tabular}{c c}
\includegraphics[]{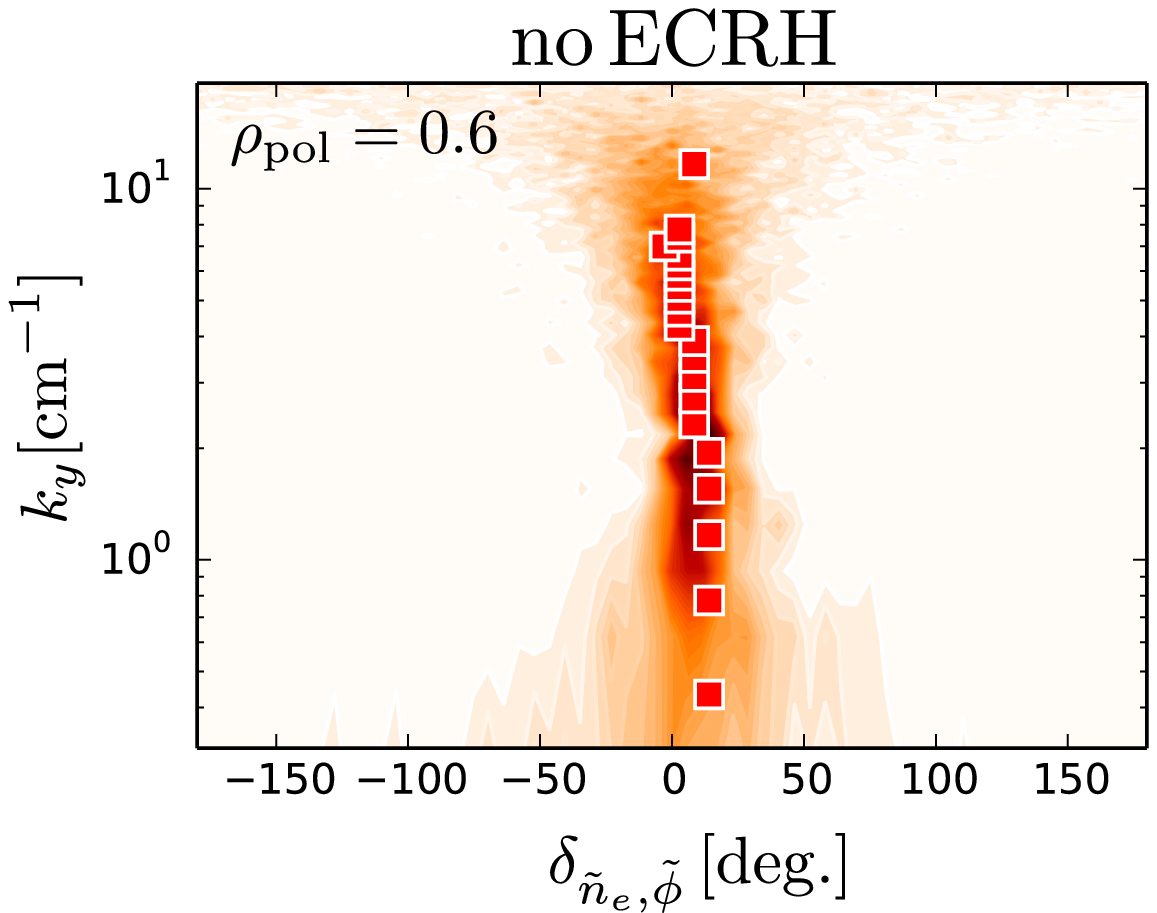} &
\includegraphics[]{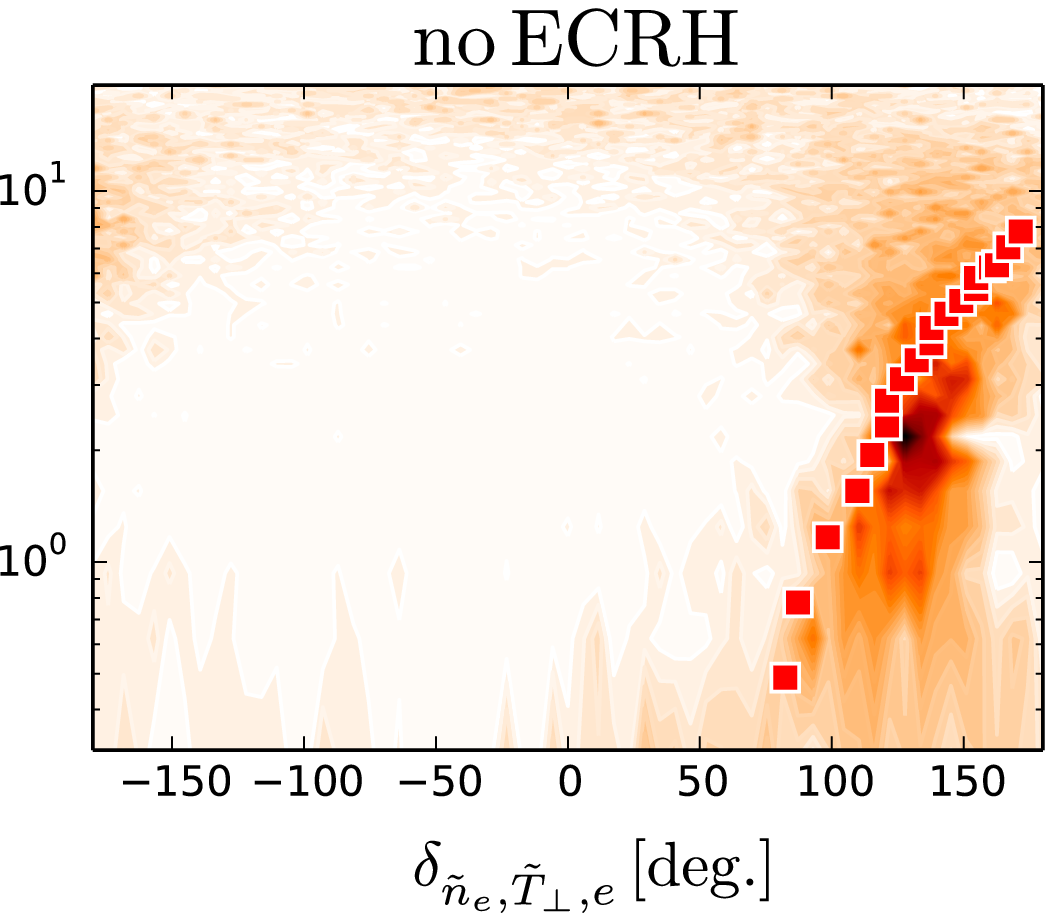} 
\end{tabular}
}
\caption{(color online) Case without ECRH at $\rho_{\rm pol} = 0.6$: Comparison of linear (markers) and nonlinear (contour) cross-phases as function of the binormal wavenumber  where amplitudes increase from white to black.  Left. Cross-phases between electron density and electrostatic potential fluctuations.  Right. Cross-phases between density and temperature fluctuations.}
\label{fig_cross}
\end{figure}

Finally, in  Fig.~\ref{fig_cross}, the colored contours display the cross-phases obtained from the nonlinear simulations, while the red squares are used to display the cross-phases of the linear simulations for  the case without ECRH at $\rho_{\rm pol} = 0.6$.   The agreement between linear and nonlinear cross-phases is remarkably good. This is also observed  for  the other cases~\cite{dannert05,told13,tobias14}. This result implies that linear simulations  could be enough  to compare with experimental results.  
%


\section{Discussion \label{discussion}}

\begin{figure}[h]
\resizebox{\columnwidth}{!}{
\begin{tabular}{c}
\includegraphics[]{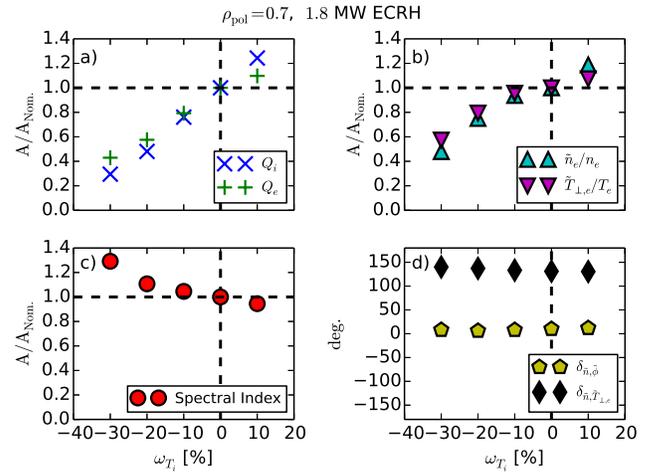} 
\end{tabular}
}
\caption{(color online)   Impact of the variation of the logarithmic ion temperature gradient around the nominal value for the case
at $\rho_{\rm pol=0.7}$ with $1.8$ MW ECRH  on  a)     ion and electron heat fluxes   b)  electron density  and temperature fluctuation amplitudes, c) electron density  spectral index and d) cross phases.
 Figures a), b) and c) are normalized with respect to  the value at the nominal ion temperature gradient.}
\label{fig_variation}
\end{figure}

Focusing on the case at   $\rho_{\rm pol} = 0.7$ with $1.8$ MW ECRH, the key results obtained in this paper 
can be illustrated in Fig.~\ref{fig_variation}. Here, we show the impact of  the variation of the logarithmic ion 
temperature gradient  around the nominal value on various observables.  
Ion and electron heat fluxes (a), electron and temperature fluctuation amplitudes (b) and  electron density spectral indices (c)  are all very sensitive with respect to small changes in the ion logarithmic temperature gradient. 
For instance,  by decreasing $\omega_{T_i}$ by $30\%$, the  amplitudes can be reduced   by a factor of $2$ for the density and temperature fluctuations and by a factor of $4$ for the heat fluxes. This result implies that the comparison of gyrokinetic simulation and experimental measurements  for these observables are very sensitive to uncertainties in the experimental input profiles. 
On the contrary, cross-phases between density and temperature fluctuations and between density and  electrostatic potential (d) are rather insensitive with respect to $\omega_{T_i}$. This, together with the fact that linear  and nonlinear cross-phases agreed also remarkably well, indicates that cross-phases could be a good  observable to compare 
(fast)  linear gyrokinetic simulations with experimental measurements. 


\section{Conclusions and future work \label{conclusions}}

We have analyzed, by means of gyrokinetic simulations with  GENE, core turbulence features
of an H-mode discharge  in ASDEX Upgrade.  The main results  of this paper can be summarized as follows.
 Flux-matched simulations were achieved by varying the nominal ion temperature gradient by a factor of $20-30 \%$, which  is within the uncertainty range of the experimental profiles. 
  In addition, density  fluctuation levels show an agreement  in  the shape of the radial turbulence  level profiles, although the effect of the  ECRH on the fluctuation levels was not reproduced.  Non-universal power-law spectra were found for  turbulence driven by  ITG modes. In particular, ITG instability  exhibits  spectral indices  for the density fluctuation spectra which cover a broad range of values. These values depend on the radial position and also on the specific  ion temperature gradient. Gyrokinetic simulations predict a decrease of the exponents with respect to both the increase of the ion temperature gradient and the increase of the radial position.  These results could help  validate future analytical theories and are useful for comparisons with other gyrokinetic codes and future measurements. 
 Regarding the electron temperature fluctuations, we observe for the inner position ($\rho_{\rm pol} = 0.6$) fluctuation amplitudes which are close to the sensitivity of the CECE diagnostic.  Therefore, it is possible that only measurements   at positions larger than   $\rho_{\rm pol}=0.6$  could be detected with this diagnostic in such  discharges. We hope that these results can provide guidance for the development of the CECE diagnostic that is currently being  installed in ASDEX Upgrade. 
Finally, by analyzing   cross-phases between density and temperature fluctuations and between density and electrostatic potential, we observed that linear and nonlinear cross-phases agree remarkably well, and that they are  rather insensitive with  respect to the variation of the ion temperature gradient,  indicating that cross-phases could be a good observable  with experimental measurements for comparisons. 

For future work, GENE and ASDEX Upgrade comparisons will continue with  the study of  similar H-modes plasmas but with higher ECRH power (up to $3.6$  MW). These discharges  are expected to have  peaked electron temperature profiles and allow TEM modes to be dominant. This will allow us to  study fundamental differences between ITG and TEM modes from  a microscopic level. Furthermore, additionally dedicated discharges  have already been conducted in  which detailed wavenumber spectra have been measured with the Doppler reflectometer. These comparisons,  along with the inclusion of  future CECE measurements,  will help in further validating gyrokinetic codes and   the development of  synthetic diagnostics.


\section*{Acknowledgements}
The authors would like to thank A.~E.~White, G.~D.~ Conway and 
U.~Stroth for fruitful discussions. The simulations presented in this work were carried out
using the HELIOS supercomputer system at the Computational
Simulation Centre of International Fusion Energy Research
Centre (IFERC-CSC), Aomori, Japan, and the HYDRA supercomputer
 at the Rechenzentrum Garching (RZG), Germany. 
 This work has been carried out within the framework of the EUROfusion Consortium 
 and has received funding from the Euratom research and training programme 2014-2018 under grant agreement No 633053. The views and opinions expressed herein do not necessarily reflect those of the European Commission.  The research leading to these
 results has also received funding from the European Research Council under the European Unions Seventh Framework Programme (FP7/2007V2013)/ERC Grant Agreement No. 277870.



\begin{thebibliography}{99.}
\bibitem{normaliz}
In Ref.~\onlinecite{tim14}, the gradients are given in terms  of the major radius in opposite to the reference length (proportional to the minor radius) used in this paper,  see  Table~\ref{table_exp} for the specific values.
\end{thebibliography}
\end{document}